\documentclass[sigplan,screen]{acmart}
\AtBeginDocument{%
  \providecommand\BibTeX{{%
    Bib\TeX}}}

\copyrightyear{2025}
\acmYear{2025}
\setcopyright{acmlicensed}
\acmConference[FSE Companion '25]{33rd ACM International Conference on the Foundations of Software Engineering}{June 23--28, 2025}{Trondheim, Norway}
\acmBooktitle{33rd ACM International Conference on the Foundations of Software Engineering (FSE Companion '25), June 23--28, 2025, Trondheim, Norway}
\acmDOI{10.1145/3696630.3728591}
\acmISBN{979-8-4007-1276-0/2025/06}

\usepackage{listings}
\usepackage{subcaption}
\usepackage{balance}
\usepackage{multirow}
\usepackage{algorithm}
\usepackage{graphicx}
\usepackage{textcomp}
\usepackage{url}

\def\BibTeX{{\rm B\kern-.05em{\sc i\kern-.025em b}\kern-.08em
    T\kern-.1667em\lower.7ex\hbox{E}\kern-.125emX}}

\newcommand\tsdetector{\textit{TS-Detector}}
\newcommand\etal{\textit{et al.}}

\definecolor{codegreen}{rgb}{0,0.6,0}
\definecolor{codegray}{rgb}{0.5,0.5,0.5}
\definecolor{codepurple}{rgb}{0.58,0,0.82}
\definecolor{backcolour}{rgb}{0.95,0.95,0.92}

\lstdefinestyle{mystyle}{
    backgroundcolor=\color{backcolour},   
    commentstyle=\color{codegreen},
    keywordstyle=\color{magenta},
    numberstyle=\tiny\color{codegray},
    stringstyle=\color{codepurple},
    basicstyle=\ttfamily\footnotesize,
    breakatwhitespace=false,         
    breaklines=true,                 
    captionpos=b,                    
    keepspaces=true,                 
    numbers=left,                    
    numbersep=5pt,                  
    showspaces=false,                
    showstringspaces=false,
    showtabs=false,                  
    tabsize=2
}

\begin{document}

%\title{Detecting Feature Toggle Usage Patterns : A Python Based Tool}

\title{\textit{\tsdetector{}} : Detecting Feature Toggle Usage Patterns}

\author{Tajmilur Rahman}
\affiliation{%
  \institution{Gannon University}
  \city{Erie}
  \state{PA}
  \country{USA}
}
\email{rahman007@gannon.edu}

\author{Mengzhe Fei}
\affiliation{%
  \institution{University of Saskatchewan}
  \city{Saskatoon}
  \state{SK}
  \country{Canada}
}
\email{mef382@usask.ca}

\author{Tushar Sharma}
\affiliation{%
  \institution{Dalhousie University}
  \city{Halifax}
  \state{NS}
  \country{Canada}
}
\email{tushar@dal.ca}

\author{Chanchal Roy}
\affiliation{%
  \institution{University of Saskatchewan}
  \city{Saskatoon}
  \state{SK}
  \country{Canada}
}
\email{chanchal.roy@usask.ca}

\begin{abstract}
%Feature toggles are conditional variables for controlling the flow of program execution. 
Feature toggles enable developers to control feature states, allowing the features to be released to a limited group of users while preserving overall software functionality. 
The absence of comprehensive best practices for feature toggle usage often results in improper implementation, causing code quality issues.
Although certain feature toggle usage patterns are prone to toggle smells, there is no tool as of today for software engineers to detect toggle usage patterns from the source code. 
This paper presents a tool \tsdetector{} to detect five different toggle usage patterns across ten open-source software projects in six different programming languages.
We conducted a manual evaluation and results show that the true positive rates of detecting Spread, Nested, and Dead toggles are 80\%, 86.4\%, and 66.6\% respectively, and the true negative rate of Mixed and Enum usages was 100\%.
The tool can be downloaded from its GitHub repository~\footnote{\tsdetector{} repository: https://github.com/tajmilur-rahman/toggle-smells} and can be used following the instructions provided there. 
The video demonstration can be viewed from the link\footnote{Video demonstration: https://doi.org/10.5281/zenodo.14086522} shared.
\end{abstract}


\maketitle

\section{Introduction}
\textit{Feature toggles}, also known as flags, switches, and flippers, are used in conditional statements to enable or disable users access to certain features~\cite{fowler2017feature}. 
Primarily, feature toggles are used for feature branching and merging in Continuous Integration and Deployment (CI/CD) environments~\cite{mahdavi2021software}. 
The streamlined development processes emphasize the need for flexible and efficient deployment strategies over iterative life cycles. 
In the current competitive business environment, feature toggles have emerged as indispensable tools, offering developers the ability to control the access to specific features to a subset of users in real-time, facilitating seamless integration, testing, and release cycles. 
However, it comes with additional complexity to the existing code and
significant responsibilities such as 
% deleting unused toggles and 
awareness of toggle life span 
% since feature toggles typically have a longer life span~\cite{rahman2016feature}. 
and managing their longer life cycle~\cite{rahman2016feature}.

In the absence of best practices or coding standards, developers often use feature toggles in ``if'' conditions. 
As feature toggles gain popularity in software companies ~\cite{mahdavi2021software}, their improper maintenance has led to notable disasters~\cite{knightcapital}, highlighting the need to identify and categorize improper usage. 
Improper implementation of feature toggles, similar to code smells, introduces maintainability issues and is referred to as toggle smells~\cite{refactoringBook, Sharma2018}. 
Despite increased academic awareness of toggle misuse~\cite{rahman2023feature, rahman2016feature}, no tool currently exists to detect usage patterns. 
Such a tool would help developers identify toggle smells and refactor them, mitigating maintainability issues.

This study presents a tool named as \tsdetector{} to detect five potential toggle smells in code written mainly in six popular languages (i.e. Java, C/C++, Python, Go, and C\#).  \
Previous study~\cite{rahman2024exploring} shows that the following usage patterns have potentials to be considered as toggle smells, therefore we develop the tool to detect:
\textit{dead}, \textit{nested},
\textit{spread}, \textit{mixed}, and \textit{enum} usage patterns. 
% in \textit{Pytorch}~\cite{pythorch}, \textit{Sentry}~\cite{sentryGithub} in Python, \textit{Cadence}~\cite{cadenceGithub}, \textit{Temporo}~\cite{temporo} in Go,  \textit{OpenSearch}~\cite{opensearchGithub}, \textit{SDB2}~\cite{sdb2} in Java, \textit{Chromium}~\cite{chromiumGithub}, and \textit{Dawn}~\cite{dawnGithub} in C and C++. 
% The selected languages are selected from the most commonly used languages in the modern industry.
The tool offers customizations in the form of commands 
% way of using the tool allowing
% analysis of software projects with diverse organization and source code structure.
to accommodate differences in programming languages and the way a development team use feature toggles.

The key contributions are, i) we reassess the algorithms from the previous study~\cite{rahman2023feature} and found that the Combinatorial usage is only available in Chromium source code as a special use case to adapt with the old command-line feature toggles, ii) we have implemented remaining five algorithms, iii) we test the tool on ten open-source projects and manually evaluate on five of them, iv) we propose some best-practices based on our experience gained during the tool development.

\section{Background \& Related work}

\begin{figure*}[ht!]
    \centering
    \includegraphics[width=0.8\textwidth]{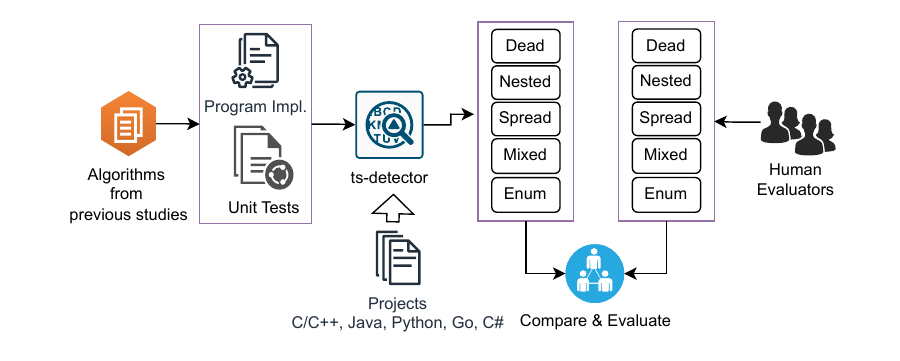}
    \caption{Development method.}
    \label{fig:methodology}
\end{figure*}

Large software companies, such as Google~\cite{chromeUseFlags}, Facebook~\cite{facebookUseFlags}, Uber~\cite{uberUsePiranha}, Netflix~\cite{netflixTechBlog}, Flickr~\cite{flickrFlippingOut}, and Apptimize~\cite{apptimizeDoc}, 
% have transitioned to more concise release models using 
use feature toggles to roll out features in a controlled manner.
Along the similar lines,
Martin Fowler~\cite{fowler2017feature} and Kim~\cite{kim2021devops} 
emphasized the significance of feature toggles in continuous deployment and feature development. 

Hezaveh \etal{}~\cite{mahdavi2021software} elucidated the utilization of feature toggles in the software development through qualitative analysis, classifying them into five distinct types: \textit{release toggles, experiment toggles, ops toggles, permission toggles}, and \textit{development toggles}. 
Their study offered comprehensive insights into operational techniques and diverse strategies for managing feature toggles.

Usage patterns of the C/C++ pre-processors i.e. \textit{``\#ifdef''}s have been studied by many researchers~\cite{liebig2011analyzing,medeiros2017discipline,le2011ifdef,fenske2015code}.
Medeiros \etal{}~\cite{medeiros2017discipline}, and Leibig \etal{}~\cite{liebig2011analyzing} discussed the disciplines of using \textit{\#ifdef}s.
Meinicke \etal{}~\cite{meinicke2020exploring} explored the differences and commonalities between feature toggles (referred to as flags) and C pre-processors. 
Our study focuses on the usage patterns of feature toggles where one of the usage patterns named as \textit{mixed usage pattern} involves both pre-processors and run-time toggle variable.

Although, many studies have been conducted on feature toggles, however, none of the existing works propose a tool support to detect toggle usage patterns for software developers to evaluate their potentiality of becoming toggle smells.
We believe our tool \tsdetector{} will enable other researchers to accelerate research in the field and assist software engineers investigate toggle usage in source code.

\section{Implementation}
\subsection{Overview}
%\todo{goal of the study}
\tsdetector{} is developed to detect feature toggle usage patterns present in source code. 
Figure~\ref{fig:methodology} offers an overview of the development of the tool.
The tool was initiated in our previous study~\cite{rahman2024exploring} supporting only C++ language and was tested on Google Chromium and Dawn. 
In this revised tool, we not only added the support to other programming languages and feature toggles
but also conduct thorough evaluation.
% We further review all six algorithms developed earlier for each six toggle usage patterns and assess the toggle usage patterns manually from ten open source projects. 
% We found that the ``Combinatorial''~\cite{rahman2023feature} usage pattern was a special usage pattern only found in Google Chromium source code. 
% The purpose of such usage pattern was to adapt the old command-line toggles with the new feature toggles. 
% This made us decide to move forward with the rest of the five usage patterns towards implementation. 
We then develop functions for five usage patterns, namely, 
\textit{dead usage}, \textit{nested usage},
\textit{spread usage}, \textit{mixed usage}, and \textit{enum usage} patterns.
The functions are written in Python that support C, C++, Python, Java, Go, and C\# language based projects. 
% We followed Test Driven Development (TDD) approach where each toggle usage detector function comes with a set of unit test methods.
% TDD is different than writing unit tests.

Once the tool is developed we test it on ten open source projects. 
We also employ a team of human developers to find toggle usage patterns based on the logic built 

\subsection{Algorithms}
Toggle usage patterns are first discussed in our first study on feature toggles~\cite{rahman2016feature}. 
We refine and extend the algorithms from our previous study~\cite{rahman2023feature, rahman2024exploring} to identify toggle usages in a reliable and generalizable manner.
Although during practical implementation previously explained algorithms have received minor corrections, due to space constraint we are not repeating all the algorithms here.
However, we would like to mention one issue regarding Spread toggles. 
If a feature toggle is used in multiple components then this type of spreading nature of usage is referred to as spread toggles~\cite{rahman2023feature}.
It could be challenging for a user to understand how the software is structured and which components the toggle variables are spreading in.
It is imperative to establish a definition for these components that strikes a balance, avoiding both excessive specificity and excessive generality, to mitigate the occurrence of false positives and false negatives in detecting spread toggles. 
Such definitions may vary across programming languages, in most cases, we choose a class to break up the code. 
However, there are languages where class constructs are absent. 
In the case of Go language, a package, a collection of files that will be compiled together, emerges as a pragmatic choice for representing code segments, owing to its inherent versatility and impartiality across diverse language contexts.

We collect two software projects for each programming language and download their source code from their official GitHub repositories\footnote{Links to the projects \tsdetector{} was tested on are provided here https://github.com/tajmilur-rahman/toggle-smells} and identify the configuration files from each of them.
Here, Chromium and Dawn represents C and C++ both languages, therefore, we had ten projects instead of twelve representing six languages.

Toggles are typically maintained in configuration files following a specific naming convention.
The toggle configuration files vary project to project in terms how the the feature toggles are configured.
\tsdetector{} takes the configuration file path as one of the inputs by the user.
%Following code snippet of Chromium shows some of the feature toggles to control GPU features specified in a toggle configuration file \textit{gpu\_switches.cc}.
%\begin{lstlisting}[language=C, caption={Chromium: GPU toggles configured in gpu\_switches.cc}]
%// Disable the GL error log limit.
%const char kDisableGLErrorLimit[] = "disable-gl-error-limit";
%// Disable the GLSL translator.
%const char kDisableGLSLTranslator[] = "disable-glsl-translator";
%// Turn off user-defined name hashing in shaders.
%const char kDisableShaderNameHashing[] = "disable-shader-name-hashing";
%// Turn on Logging GPU commands.
%const char kEnableGPUCommandLogging[] = "enable-gpu-command-logging";
%\end{lstlisting}
%
%The project specific toggles not only are limited to the specific way of naming them and declaring them in their own way;
%the structure of toggle variable specification also tends to be project-specific.
%For example, in a large project like Chromium, each component has its own set of feature toggles configured in a separate file with a similar naming convention of \textit{*\_switches.cc}. 
%On the other hand, Dawn lists all feature toggle variables in one single file.
%To mitigate this issue our tool allows users to provide the configuration file name in a wildcard fashion. Meaning, users can provide the full name if it is the only configuration file for an exact match(Example: ``Features.cpp''), or a wildcard name for a broader match(Example: ``*\_switches.*''). 

\textbf{Input parameters:} \tsdetector{} supports four input parameters as listed below. Two of them are required and two others are optional. 
The tool can be used without installing it in the local machine.
\begin{lstlisting}[language=HTML, caption={Parameters}]
SYNOPSIS:
python3 tsd.py -p [options] -c [options] ... [options]

DESCRIPTION:
-p Project source code. Required.
-c Toggle configuration file. Required.
-o Output file. Optional.
-t Toggle usage pattern. Optional.
-l Language of the project. Optional.
\end{lstlisting}

Users can use the following command and parameters to use \tsdetector.
\begin{lstlisting}[language=Python, caption={Command template to use ts-detector}]
python3 tsd.py -p </source/path> -c </config/path> -o <output/path> -t pattern
\end{lstlisting}

\textbf{Output:} The output of the tool is a JSON object consists of toggle usage pattern, and number of instances found. 
%Following is an example output from Sentry project.
%
%\begin{lstlisting}[language=C, caption={Output JSON of ts-detector}]
%{
%  "dead": {
%    "toggles": [], "qty": 0
%  },
%  "spread": {
%    "toggles": [
%      "VSTEST_DISABLE_MULTI_TFM_RUN",
%      ...
%      "VSTEST_DISABLE_UTF8_CONSOLE_ENCODING"
%    ], "qty": 5
%  },
%  "nested": {
%    "toggles": [
%      "VSTEST_DISABLE_STANDARD_OUTPUT_CAPTURING",
%      ...
%      "VSTEST_DISABLE_UTF8_CONSOLE_ENCODING"
%    ], "qty": 8
%  },
%  "mixed": {
%    "toggles": [], "qty": 0
%  },
%  "enum": {
%    "toggles": [], "qty": 0
%  }
%}
%\end{lstlisting}

\begin{figure*}[ht!]
    \centering
    \includegraphics[width=0.8\textwidth]{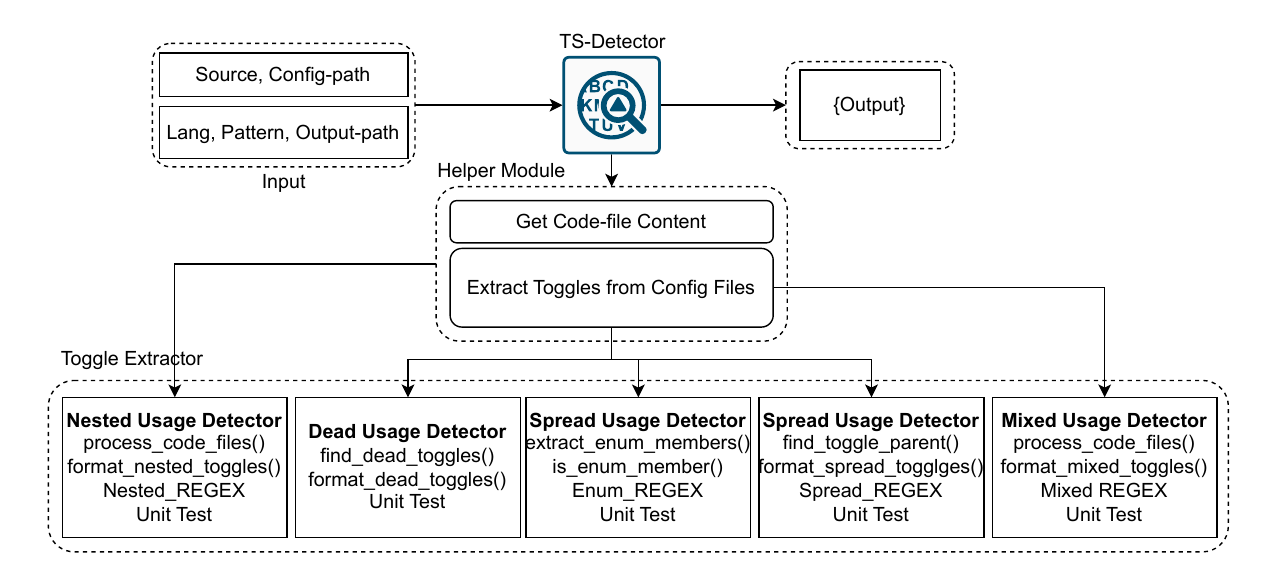}
    \caption{\tsdetector{} Architecture.}
    \label{fig:arch}
\end{figure*}

Figure~\ref{fig:arch} shows the architecture of \tsdetector{} 
% We design generic detection strategies for each toggle pattern using regular expressions.
% To make sure that we identify all toggle patterns in C++ projects,
% % are generalized among C++ projects and not specific to Chromium and Dawn only, 
% we design our algorithms to take a configuration file as one of the parameters.
When the user provides input, the core \tsdetector{} module manages the overall workflow by interacting with a helper module. 
The helper module extracts content from code files and identifies toggles from configuration files through utility functions by matches all variables with a set of predefined general regular expressions. 
It further filters the toggle variables based on rules such as duplicates, name length, and specific language keywords. 
These helper functions prepare the toggle data to facilitate the detection process.

The logic is divided into specialized detectors for each toggle smell, using language-specific regex for toggle patterns. The Dead Usage detector searches for toggles across files without regex. The Spread Usage detector identifies toggle contexts based on language rules (e.g., Golang packages, C\# namespaces). The Mixed Usage detector targets restricted patterns in C and C++.
The Nested usage detector is more complex needing to identify toggles within nested structures
%such as \texttt{if}/\texttt{elif} statements, \texttt{return} statements, 
and variable assignments. 
%In some cases, other variables are created representing the toggle and used within nested constructs, which must also be detected. 
The Enum Toggle Detector is less frequently used but identifies toggles that match the enum pattern specific to each language. 
%Finally, the tool generates the output in the desired format summarizing the detected toggles.

\section{Evaluation}
The \tsdetector{} has been tested on ten open source software projects. We have manually evaluated the tool on five of them representing six programming languages.
Meaning, we manually identified all five usage patterns and compare that manual finding with what \tsdetector{} found. 
The evaluation is presented by True Positive (TP), False Positive (FP), and False Negative (FN).
If a toggle usage is found both manually and by the tool, then it's a TP, if a toggle usage is not found manually but found by the tool then it's an FP, and if a toggle usage is found manually but not by the tool, then it's a FN. 

\begin{table}[htbp]
\caption{Evaluation of \tsdetector{}}
\begin{center}
\label{tab:manual-eval}
\begin{tabular}{|l|l|l|l|l|l|}
\hline
{} & Manual & \tsdetector & TP & FP & FN \\
\hline
\multicolumn{6}{|c|}{\textbf{Spread}} \\
\hline
OpenSearch & 6 & 5 & 4 & 1 & 2 \\
\hline
Temporal & 314 & 301 & 299 & 2 & 15 \\
\hline
Dawn & 123 & 129 & 116 & 13 & 4 \\
\hline
Sentry & 146 & 200 & 124 & 79 & 22 \\
\hline
Server & 9 & 16 & 6 & 10 & 3 \\
\hline
\multicolumn{6}{|c|}{\textbf{Nested}} \\
\hline
OpenSearch & 2 & 2 & 2 & 0 & 0 \\
\hline
Temporal & 8 & 8 & 8 & 0 & 0 \\
\hline
Dawn & 62 & 70 & 52 & 18 & 10 \\
\hline
Sentry & 86 & 81 & 70 & 11 & 16 \\
\hline
Server & 2 & 13 & 2 & 11 & 0 \\
\hline
\multicolumn{6}{|c|}{\textbf{Dead}} \\
\hline
OpenSearch & 1 & 3 & 1 & 2 & 0 \\
\hline
Temporal & 2 & 3 & 2 & 1 & 0 \\
\hline
Dawn & 3 & 8 & 3 & 5 & 0 \\
\hline
Sentry & 113 & 131 & 97 & 134 & 16 \\
\hline
Server & 27 & 23 & 23 & 0 & 4 \\
\hline
\end{tabular}
\end{center}
\end{table}

In this section we present and discuss the evaluation results. Table~\ref{tab:manual-eval} shows the number of toggle usage patterns found manually and by the tool along with TP, FP, and FN.  
We selected the projects for manual detection randomly, regardless of any special consideration. 
Although, Google Chromium has a few Mixed, and Enum usage patterns, none of the projects that we manually checked showed any presence of such usage patterns. 
Temporal, Dawn, and Sentry are relatively larger projects than the two others OpenSearch and Server, as a result, the toggle usages in these projects are higher in number.

\begin{figure}[ht!]
    \centering
    \includegraphics[width=\columnwidth]{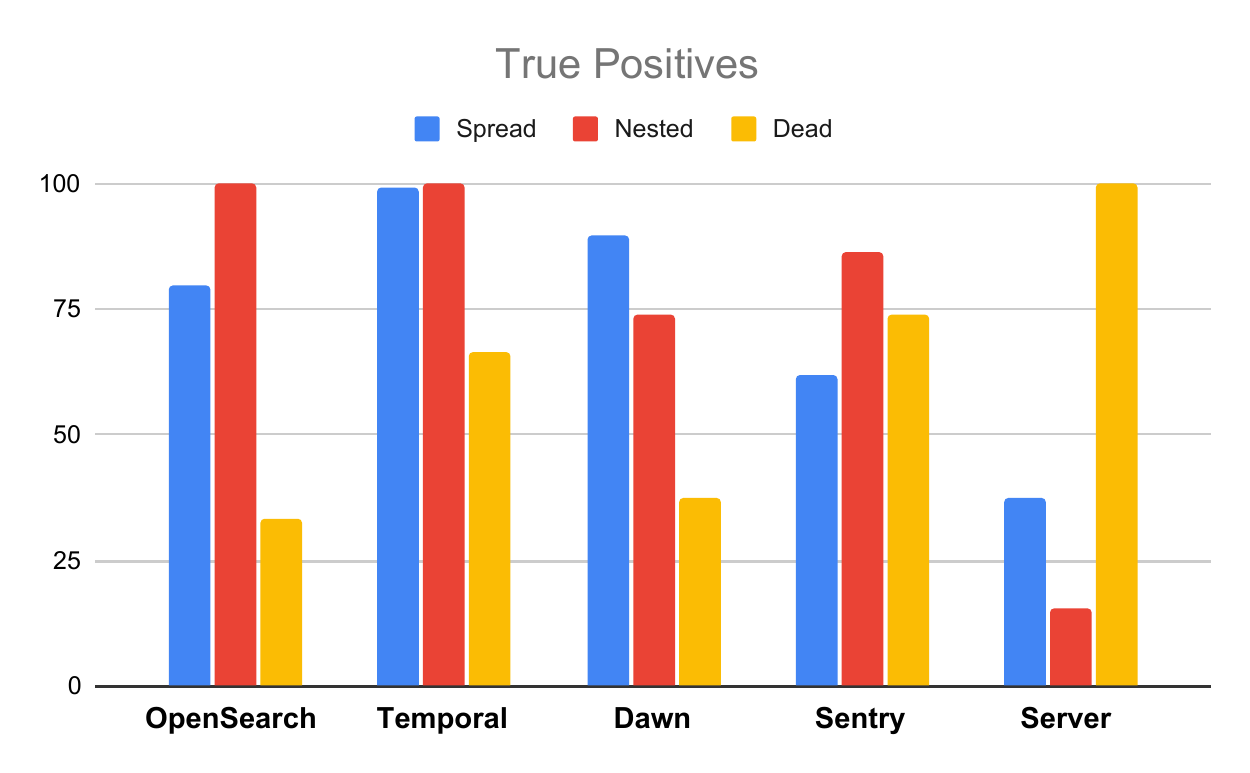}
    \caption{True-Positives of detecting toggle smells using \tsdetector{}.}
    \label{fig:tp}
\end{figure}

Figure~\ref{fig:tp} shows the percentages of true positives for each project in identifying Spread, Nested, and Dead toggles. 
In OpenSearch 6 spread usages were found manually while the tool identified 5 with 4
true positives (80\%), and 1 false positive. 
One additional usage was identified (false negative FN) by the tool which was not actually a toggle usage and wasn't identified manually.
In Temporal, and Dawn TP for the spread usages were 299, and 116 with a very high percentage of true positives (TP 99.3\% and 89.9\%) respectively. 
In Sentry TP was higher than FP (TP 62\% FP 37.5\%). However, Sentry and Server have higher FP compared to the other projects.
This is because in these two projects toggle variables are configured in the same configuration file as many other global variables and project-wide constants. 
Moreover, the regular globals and constants are syntactically similar to the toggle variables. 
As a result, our tool has identified many of those constants and gobals as feature toggle variables. 

Similar results observed for nested usages too. 
OpenSearch, Temporal, Dawn, and Sentry have high true positives of 100\%, 100\%, 74.2\%, and 86.4\% respectively with only exception of Server (TP 15.4\%) with a very high 84.6\% of false positives. 
Only two nested usages were found manually for Server while 11 other global variables in the toggle configuration file were identified as toggle variables by \tsdetector{} and they were in nested usage pattern throughout the source code.

Dead usage identification is lower than spread and nested usage for OpenSearch, Temporal, and Dawn. In OpenSearch, the tool identified 3 dead toggles, but only 1 was truly dead; inconsistent access to toggle variables caused misidentification. Similarly, in Temporal, the tool found 3 dead toggles instead of 2 due to confusion with a non-toggle variable in the configuration file. The true positive rates for Sentry and Server were 74\% and 100\%, respectively.

\balance{}
\section{Conclusions and Future Work}
%\tsdetector{} is the first tool that detects toggle usage patterns. 
%It still has plenty of rooms of improvements. 
%%Observations and Suggestions
%We observed that having both regular constants, global variables, and the feature toggle variable together in the same configuration file makes the toggle identification challenging. 
%This makes detecting and tracking feature toggles imperfect which and becomes an obstacle to automatically detecting toggle smells.
%Previously the algorithms were developed based on C/C++ language based projects only. 
%In this study we achieved a coverage of six programming languages generalizing the toggle usage detection algorithms cross language and cross projects. 
%We encountered a lot of challenges and learned while developing this tool.
%It is highly important to configure toggle variables properly for enhanced maintainability and better traceability. Based on our experience we propose the following best practices of configuring and using feature toggles: i) Avoid regular global variable and toggle variables in config files. ii) Toggle variable names should be unique not a substring of another toggle variable. iii) Nesting should not occur more than one level. iv) Although spreading is natural since the features crosscut components, it should not spread among unrelated components. v) Access to toggle variables should be consistent.
`\tsdetector{}' is the first tool to detect toggle usage patterns, but it has room for improvement. Mixing regular constants, global variables, and toggle variables in the same configuration file complicates toggle identification, making automatic detection of toggle smells imperfect. The tool now supports six programming languages, generalizing detection algorithms across languages and projects. 
%Based on our experience, we propose best practices for feature toggles:  
%1. Avoid mixing global and toggle variables in config files.  
%2. Use unique toggle variable names, avoiding substrings of others.  
%3. Limit nesting to one level.  
%4. Minimize spreading toggles across unrelated components.  
%5. Ensure consistent access to toggle variables.  
Our future work will continue improving the accuracy, expanding usability, and adding a user interface for advanced and comfortable user interaction.


\bibliographystyle{ACM-Reference-Format}
\bibliography{main}

\end{document}